\documentclass[%
preprint,
superscriptaddress,
showpacs,preprintnumbers,
amsmath,amssymb,
aps,
prb,
]{revtex4-1}

\usepackage{sidecap}
\sidecaptionvpos{figure}{c}

\usepackage{graphicx}
\usepackage{dcolumn}
\usepackage{bm}
\usepackage[
colorlinks=true,
linkcolor=blue,
citecolor=blue,
filecolor=blue,
urlcolor=blue]{hyperref}

\usepackage{amsmath,amsfonts,amsgen,amsbsy,amsopn,amstext,amssymb}
\usepackage{epstopdf}
\usepackage{color}

\usepackage[english]{babel}
\usepackage{color}
\usepackage{lineno}
\modulolinenumbers[5]

\renewcommand{\imath}{i}

\begin{document}
	
\preprint{}

\title{Formation of high-order acoustic Bessel beams by spiral diffraction gratings}

\author{No\'e~Jim\'enez}
\email{noe.jimenez@univ-lemans.fr}
\affiliation{LUNAM Universit\'e, Universit\'e du Maine, CNRS, LAUM UMR 6613, Av. O. Messiaen, 72085 Le Mans, France}
\author{R.~Pic\'o}
\affiliation{IGIC, Universitat Polit\`ecnica de Val\`encia, Paranfimf , E-46730 Grao de Gandia, Spain}
\author{V.~S\'anchez-Morcillo}
\affiliation{IGIC, Universitat Polit\`ecnica de Val\`encia, Paranfimf , E-46730 Grao de Gandia, Spain}
\author{V.~Romero-Garc\'ia}
\affiliation{LUNAM Universit\'e, Universit\'e du Maine, CNRS, LAUM UMR 6613, Av. O. Messiaen, 72085 Le Mans, France}
\author{L.~M.~Garc\'ia-Raffi}
\affiliation{IUMPA, Universitat Polit\`ecnica de Val\`encia, Camino de Vera s/n, 46022, Val\`encia, Spain}
\author{K.~Staliunas}
\affiliation{ICREA, Departament de F\'i­sica i Enginyeria Nuclear, Universitat Polit\`ecnica de Catalunya, Colom 11, E-08222 Terrasa, Barcelona, Spain}

\date{\today}

\begin{abstract}
	The formation of high-order Bessel beams by a passive acoustic device consisting of an Archimedes' spiral diffraction grating is theoretically, numerically and experimentally reported in this work. These beams are propagation-invariant solutions of the Helmholtz equation and are characterized by an azimuthal variation of the phase along its annular spectrum producing an acoustic vortex in the near field. In our system, the scattering of plane acoustic waves by the spiral grating leads to the formation of the acoustic vortex with zero pressure on-axis and the angular phase dislocations characterized by the  spiral geometry. The order of the generated Bessel beam and, as a consequence, the size of the generated vortex can be fixed by the number of arms in the spiral diffraction grating. The obtained results allow to obtain Bessel beams with controllable vorticity by a passive device, which has potential applications in low-cost acoustic tweezers and acoustic radiation force devices.
\end{abstract}

\maketitle



\section{Introduction}


Diffraction of waves, the spreading of the wave packet upon propagation, represents one of the most known properties of the wave physics. However, there are particular solutions of the wave equation that are immune to diffraction. Among them, the Bessel beams \cite{durnin1987} are diffraction free solutions with remarkable features. Of special interest are the High-Order Bessel Beams (HOBBs), characterized by high amplitude concentric rings with a profile given by the $n$th-order Bessel function in the plane transverse to the beam axis. Any Bessel beam is characterized by an annular radiation in the far field, therefore the magnitude of the spatial spectrum of these beams does not depend qualitatively on the order of the beam. The difference between zeroth and HOBBs is that the phase of the HOBBs shows a linear variation along its annular spectrum in the azimuthal direction. Thus, the wave field presents a helicoid phase dependence containing screw-type phase singularities, leading to an intensity minima at the beam axis. Solutions of such kind are of infinite transverse extent and thus can not be generated experimentally. However, it is possible to generate finite size approximations to Bessel beams which propagate over extended distances in a diffraction free manner providing potential applications for the wave physics community\cite{vasara1989, arlt2000, Oemrawsingh2004,marston2006, mitri2009,mitri2009b,jimenez2014}. While zeroth-order Bessel beams present a bright central maximum and can be useful for applications that requires focusing of energy\cite{jimenez2015}, the vortex beams generated by the HOBBs can be useful for manipulating particles in both optics \cite{pfeiffer2015} and acoustics\cite{Baresch16}. 

In the case of electromagnetic (optical) waves, vortex beams have been experimentally demonstrated by means of computer generated holograms \cite{vasara1989,heckenberg1992} or by axicons illuminated with a Laguerre-Gaussian mode \cite{arlt2000}. Other methods include an azimuth-dependent retardation on the optical field using Spiral Phase Plates (SPP) \cite{Oemrawsingh2004}, or diffraction gratings with groove bifurcation. In the latter case, vortex beams with an arbitrary topological charge have been created \cite{bekshaev2008}. This kind of beams has been shown very useful for the optical manipulation of particles. Since the first observations of manipulation of particles using optical beams \cite{ashkin1970, ashkin1987, ashkin1986}, an unexpected radial force field appeared, called gradient force, that dragged colloidal particles towards the axis in addition to the axial radiation pressure that pulls particles towards the beam. In this way, particles can be trapped into the beam axis under conditions where the dragging gradient force dominates over the pulling radiation pressure. This regime can be achieved using strongly focused light beams and for particles smaller than the wavelength, leading to possibility of manipulation of objects as small as 5 nm \cite{grier2003}. These so called optical tweezers have been employed in many other macromolecular, biological and medical applications \cite{grier2003}. However, high intensity beams are necessary for exerting strong forces leading to unwanted effects as heating, so in practical applications optical tweezers can exert forces up to 100 pN \cite{grier2003}.

Compared to optical manipulation, ultrasonic waves become advantageous to manipulate heavier objects: the smallness of the sound speed lead to larger drag forces, from 3 to 4 orders of magnitude \cite{Baresch16}, and because of the size of the acoustic wavelength bigger objects can be trapped. In addition, the interaction of Bessel beams with particles have been also intensively studied in acoustic \cite{mitri2009,mitri2009b}. Two main remarkable effects have been reported: first, the transference of orbital momentum from the acoustical vortex to the particle \cite{mitri2012,hong2015}, and on the other hand, the appearance of negative axial acoustic radiation forces \cite{marston2006,marston2007,mitri2008,mitri2009b}. This fact has motivated the development of experimental approaches to generate acoustical vortex beams. Different methods using either single acoustic sources have been developed. Phase dislocations using a single source were first proposed by Nye and Berry \cite{nye1974}. The acoustic analog of the optical SPP has been also proposed \cite{hefner1999}, consisting of a transducer with a surface properly deformed to create the helicoidal beam. This method is restricted to a single operating frequency. Other approaches with single sources, include the use of photoacoustic effect to generate an helical beam \cite{gspan2004}.

On the other hand, the generation of acoustical vortices with arrays of sources is also possible \cite{thomas2003} and have been widely used in acoustic for multiple applications: particle manipulation \cite{demore2011}, acoustical tweezers \cite{wu1991,wang2015,marzo2015,Baresch16}, angular momentum transfer \cite{volke2008}, acoustic spanners \cite{skeldon2008}, multiple-particle trapping \cite{yoon2014}, precise manipulation and sorting of cells for life sciences research \cite{li2015,guo2016} or micro-bubble capturing \cite{raiton2012}. Recently Baresch and co-workers developed the first all-acoustical single-beam trapping \cite{Baresch16}, where a negative gradient pulling force with acoustic waves was demonstrated using a single vortex beam.  However, although the array of sources provides active steering and control of the vortex beam, in such active systems the resolution of the vortex is restricted by the number of transducers in the array \cite{brunet2009}, leading to technologically complex systems in the case of vortices of high topological charge.

In this work we study the diffraction of a plane wave by a multiple-arm Archimedes' spiral diffraction grating, and propose a passive and robust method for the formation of HOBBs using such gratings. The scattering of plane acoustic waves by the spiral grating leads to the formation of the acoustic vortex with zero pressure on-axis and the angular phase dislocations characterized by the spiral geometry. The order of the generated Bessel beam and, as a consequence, the size (width) of the generated vortex can be fixed by the number of arms of the spiral diffraction grating. The obtained results allow to obtain Bessel beams with controllable vorticity by a passive device, which has potential applications in low-cost acoustic tweezers and acoustic radiation force devices. First, in Section \ref{sec:theory} we present a theoretical model for the diffraction of plane waves by the multiple-arm spiral grating with infinite radial extent. Then we numerically analyze the effects of the finite size of the sample, considering also the effects of the vibration of the scatterers. The numerical confirmation of the HOBBs is reported in Section \ref{sec:HOBB} showing the generation of the vortex in the axis and its dependence on the topological charge of the spiral diffraction grating. Finally, in Section \ref{sec:exp} we experimentally test the main results of this work by measuring the acoustic field scattered by a steel grating embedded in water. Particularly, we show the phase dislocation and the acoustic vortex generation by a first order Bessel beam.

\section{Diffraction by a spiral grating}\label{sec:theory}

\begin{figure}[t]
	\centering
	\includegraphics[width=8cm]{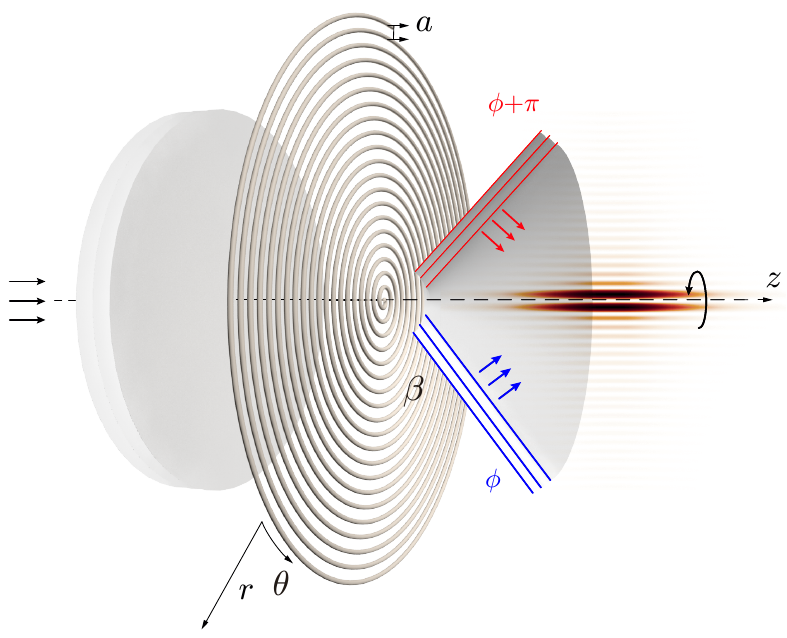}
	\caption{Scheme of the first-order Bessel beam formation by the Archimedes' spiral grating. The incident plane wave is scattered into a converging and diverging conical wavefront, in which the phase of the wave ($\phi$) is a linear function of the angle ($\theta$). A corkscrew dislocation in wavefront is produced, leading to an acoustic vortex and a zero pressure on axis, $r=0$. The minimum along the symmetry axis results from the destructive interference of first order diffraction generated at opposite sides of the axis. }
	\label{fig:Scheme} 
\end{figure}

The proposed structure is a multiple-arm Archimedes' spiral diffraction grating as shown in Fig. \ref{fig:Scheme} and \ref{fig:Figarms}. As the arms of the Archimedes' spiral present an uniform separation, the incident field is diffracted at an angle, given by diffraction grating theory. Therefore, the diffracted field is of conical wavefront as in Ref.~\cite{jimenez2014}, but here with azimuthal rotating phase due to the spiral geometry. When converging to the axis, the conical wavefront forms a HOBB. The diffracted pressure field by the grating generates an acoustic vortex line with a characteristic screw dislocation.

\subsection{Infinite diffraction gratting}

The harmonic pressure field diffracted by a grating, with source velocity distribution $v_z({\bf{r}}_0)$, can be calculated using the Rayleigh-Sommerfeld integral at any point, in cylindrical coordinates ${\bf{r}}=(r,\theta,z)$ as
\begin{equation}\label{rayleigh}
	p({\bf{r}})=\frac{- \imath \omega \rho_0 }{2 \pi} \int_{S_0} v_z({\bf{r}}_0)\frac{\exp\left(\imath k \left|{\bf{r}}-{\bf{r}}_0\right|\right)}{\left|{\bf{r}}-{\bf{r}}_0\right|}dS({\bf{r}}_0),
\end{equation}

\noindent where ${\bf{r}}_0=(r_0,\theta_0,z_0)$ is the radius vector of a surface element $dS$, $\omega$ the angular frequency, the wavenumber $k=\omega/c_0$ and $\rho_0$ and $c_0$ the density and speed of sound of the medium. The plane of the source is assumed to be at the origin of the coordinates for simplicity, $z_0=0$, then, without loss of generality we can write $v_z(r_0,\theta_0,z_0=0)=v_z(r_0,\theta_0)$. 

The Fresnel approximations assumes that the first two terms of the square root Taylor expansion are sufficient to correctly represent the phase, provided that $z$ is large enough (parabolic expansion):
\begin{equation}
\begin{split}\label{approx1}
	\left|{\bf{r}}-{\bf{r}}_0\right|=&z\sqrt{1+\frac{(x-x_0)^2}{z^2}+\frac{(y-y_0)^2}{z^2}}\simeq \\
	&z\left(1+\frac{r^2}{2z^2}+\frac{r_0^2}{2z^2}-\frac{r r_0 \cos(\theta_0-\theta)}{z^2}\right),
\end{split}
\end{equation}

\noindent and by neglecting the radial contributions in the denominator of the Rayleigh integral: $\left|{\bf{r}}-{\bf{r}}_0\right| \simeq z$.

Expressing the surface element in cylindrical coordinates $dS=r_0  d \theta_0 d r_0$, Eq.~(\ref{rayleigh}) transformes to the following form
\begin{equation}
\begin{split}\label{inte}
	 p(r,\theta,z)=&A(r,z) \int_0^\infty \int_0^{2\pi}  v_z(r_0,\theta_0) \exp\left[\imath \frac{k}{2z}r_0^2\right] \times\\
	 &\exp\left[-\imath \frac{k r } {z}r_0 \cos(\theta_0-\theta)\right]r_0  d r_0 d \theta_0 ,
\end{split}
\end{equation}

\noindent where $A(r,z)$ is independent expression on the integration variables $r_0$ and $\theta_0$:
\begin{equation}\label{amplitude}
	A(r,z)=\frac{-\imath \omega \rho_0}{2\pi z}\exp \left[\imath k \left(z+\frac{r^2}{2 z}\right)\right].
\end{equation}

\begin{figure}[t]
	\centering
	\includegraphics[width=8cm]{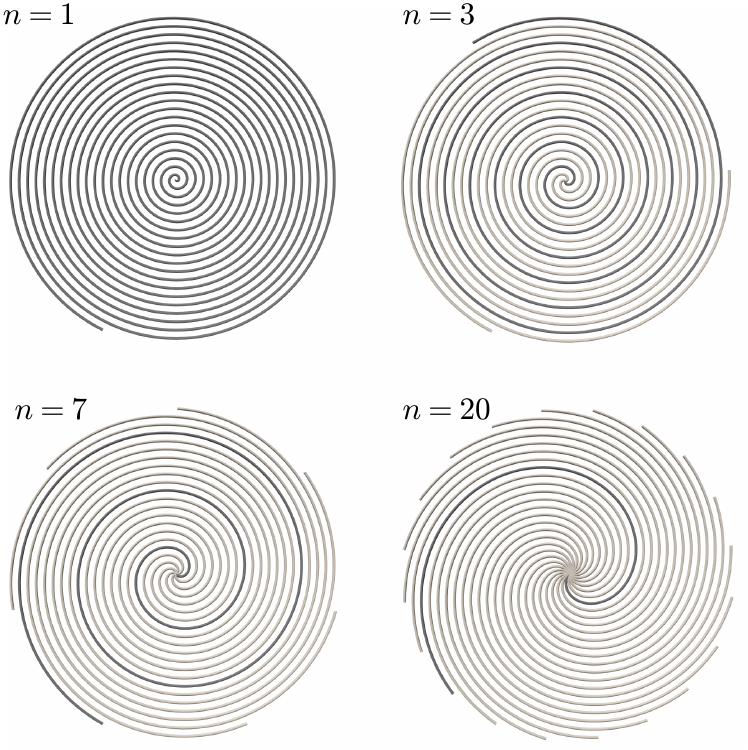}
	\caption{Examples of Archimedes' spirals with multiple arms (a) $n=1$, (b) $n=3$, (c) $n=7$ and (d) $n=20$.}
	\label{fig:Figarms}
\end{figure}

We suppose that an incoming plane wave uniformly illuminates the spiral grating at $z=0$. 
The source velocity distribution can be characterized by a complex-amplitude transmission function \cite{vasara1989}, 
\begin{equation}\label{eq:source}
	v_z(r_0,\theta_0)=v_0 \exp[-\imath k_r r_0] \exp\left[-\imath \phi(\theta_0) \right], 
\end{equation}

\noindent where $v_0$ is the particle velocity amplitude. The factor $\exp[-\imath k_r r_0]$ is the phase of the conical wavefront. The continuity of the transversal component of the wave vector $k_r$ at the interface between the homogeneous medium and the diffraction grating with scatterers separated by a distance $a$ results in $k^2=\sqrt{k_r^2+k_z^2}$ with $k_r=2\pi N/a$ and $k_z=k_r/\tan\beta$ the axial and longitudinal wavenumber, respectively, $N$ is the diffraction order and $\beta=\arcsin N \lambda / a$ is the angle of the conical wavefront with respect to the axis $z$. Here and below we will work in the range of frequencies that only excites the first diffraction order, therefore we assume $N=1$. The last factor $\exp[-\imath \phi(\theta)]$ is a phase accounting for the azimuthal dependence of the phase of the conical wavefront. 

In the case of a pure axisymmetric grating \cite{jimenez2014} where the sources are distributed in concentric circles separated at a distance $a$,
\begin{equation}\label{eq: radial}
\begin{split}
	\phi(\theta_0) = k_{r} R(\theta_0)= \frac{2 \pi}{a} R(\theta_0) &= \mathrm{const.}\,, \\ 
	\exp[-\imath \phi(\theta_0) ] &= \mathrm{cte}\,.
\end{split}
\end{equation}

\noindent Therefore, due to constant radius of each of concentric rings, the phase is independent on azimuthal angles and no vortex can be produced. In this work we consider an Archimedes' spiral grating which provides the azimuthal dependence of the phase in our system. 
The general mathematical expression for a curve describing $n$ arms of Archimedes' spirals starting from an origin can be expressed in polar coordinates as
\begin{equation}\label{spiral1}
	R( \theta_0 )=\frac{n a}{2 \pi}\theta_0  + l a,
\end{equation}

\noindent with $0\leq l \leq n-1$ the index of the $l$-th arm, and $a$ the raidal separation between arms. Figure~\ref{fig:Figarms} shows examples of spiral with multiple arms. The phase term of the Eq.~(\ref{eq:source}) can be expressed as:
\begin{equation}\label{eq:velocity1}
\begin{split}
	\exp[-\imath \phi(\theta_0)]=\exp[-\imath k_r·R(\theta_0)]=\exp \left[-\imath n\theta_0\right].
\end{split}
\end{equation}

\noindent The velocity function can be obtained by substituting Eq.~(\ref{eq:velocity1}) into Eq.~(\ref{eq:source}):
\begin{equation}\label{velocity1}
\begin{split}
	v_z(r_0,\theta_0)=v_0 \exp[-\imath k_r·r_0]·\exp[-\imath n\theta_0].
\end{split}
\end{equation}

\noindent Therefore, the particle velocity field at the source plane corresponds to a conical wavefront with an azimuthal phase rotation proportional to the number of arms of the spiral. Explicitly, the phase of the field scattered by $N$-arm spiral rotates by $2\pi N$, thus forming the phase singularity of $N$-th order.

The pressure field can be obtained by substituting the source field velocity in the double integral in Eq.~(\ref{inte}),
\begin{equation}\label{velointe}
\begin{split}
	p(r,\theta,z)=&A(r,z)  \int_0^\infty r_0 \exp \left[\left(\frac{k}{2z}r_0^2\right)\right] \times\\ 
	&\int_0^{2\pi} v_0 \exp[-\imath k_r r_0] \exp[-\imath n\theta_0]\times\\
	&\exp\left[\imath \left(\frac{k r } {z}r_0 \cos(\theta_0-\theta)\right)\right] d r_0 d \theta_0 . 
\end{split}
\end{equation}

\noindent  The terms without azimuthal dependence can be factorized out of the azimuthal integral in Eq. (\ref{velointe}) is:
\begin{equation}\label{integ}
\begin{split}
	&p(r,\theta,z)=A(r,z) v_0 \int_0^\infty r_0 \exp\left[\imath\left(\frac{k}{2z}r_0^2 - k_r r_0\right)\right] \\
	&\int_0^{2\pi} \exp[- \imath n\theta_0]\exp  \left[-\imath \frac{ k r } {z} r_0 \cos(\theta_0-\theta)\right] d r_0 d \theta_0 .
\end{split}
\end{equation}

\noindent Using the Jacobi-Anger expansion, 
\begin{equation}\label{Jacobi-expression}
\begin{split}
	J_n(\alpha)=\frac{\imath^n}{2\pi} \int_{0}^{2\pi} \exp[in\beta]\exp[-\imath \alpha \cos(\beta)] d\beta, 
\end{split}
\end{equation}
\noindent and simple algebra with a change of variable, the integration over the azimuthal angle $\theta_0$ in Eq. (\ref{integ}) can be solved and leads to:
\begin{equation}\label{Jacobi}
	p(r,\theta,z)= B(r,z) F(r,z),
\end{equation}
\noindent where 
\begin{equation}\label{Brz}
	B(r,z)=A(r,z)  2\pi v_0 \exp \left[\imath n \left(\theta-\frac{\pi}{2}\right)\right],
\end{equation}
\noindent and
\begin{equation}\label{intebessel}
	F(r,z)=\int_0^\infty r_0 \exp \left[\imath \left(\frac{k}{2z}r_0^2 - k_r r_0\right)\right] J_n\left(\frac{k r }{z}r_0\right) d r_0  .
\end{equation}
\noindent note that $n$ is the number of arms in the spiral is also the order of the Bessel function. 

The radial integral in Eq. (\ref{intebessel}) can be approximately solved by using the method of the stationary phase \cite{vasara1989,arlt2000,paterson1996}. Rapid oscillations of the exponential term of the integral in Eq. (\ref{intebessel}) mean that $F(r,z)\simeq0$ over those regions and only significant non-zero contributions to the integral occur in regions of the integration range where phase term is constant i.e., at points of stationary phase. In our case, the approximated solution to leading order of the radial integral at point $(r,\theta,z)$ reads as
\begin{equation}\label{intesolu}
\begin{split}
	F(r,z)\simeq\frac{k_r z }{k} \exp\left[-\imath \left(\frac{k r^2}{2z}+\frac{zk_r^2}{2k}\right)\right]\sqrt{\frac{2\pi z}{k}}J_n(k_r r).
\end{split}
\end{equation}
Higher order terms not considered in this solution give corrections to off-axis areas \cite{paterson1996}.

By substituting Eq. (\ref{intesolu}) into Eq. (\ref{Brz}) and Eq. (\ref{Jacobi-expression}), as well as using Eq. (\ref{amplitude}), the pressure field is written as
\begin{equation}\label{pressio}
\begin{split}
	p({\bf{r}})\simeq&-\imath p_0 k_r \sqrt{\frac{2 \pi z}{k}}\, J_n(k_r r) \exp\left[\imath k_z z  \right] \exp \left[\imath n \left(\theta -\pi/2\right)\right],
\end{split}
\end{equation}

\noindent where $p_0=\rho_0 c_0 v_0 $, and the paraxial approximation of the axial wavenumber $k_z=k \left(1  -{ k_r^2}/{2k^2} \right)$ was used. The radial distribution is given by the $n$th-order Bessel function, while the amplitude is proportional to $\sqrt{z}$, which is in fact the expression for a $n$-th order Bessel beam.  As an example we evaluate the amplitude of the pressure field along the first lobe of the first order Bessel beam ($n=1$) generated by an infinite spiral with $a/\lambda=1.2$ embedded in water. Black dashed line in Fig. \ref{fig:all4x}(a) shows the evaluation of the Eq. (\ref{pressio}) for this spiral, showing the $\sqrt{z}$ dependence. Notice that the normalized intensity, $p\;p^*/\rho_0c_0$, grows linearly with the distance $z$, with a rate given by $2\pi k_r^2J_n^2(k_rr)/k$. A simple physical interpretation of the $\sqrt(z)$ dependence in Eq. (\ref{pressio}) is that the radiation at increasing $z$ arrives scattered from the arms of spirals of increasing radius with proportionally increasing energy.

\subsection{Finite size effects}

Previous Section deals with an infinitely extended diffraction grating. This is not the real situation in experiments different finite size effects can be present. In order to analyze these finite size effects we have applied two different methods. On one hand we have numerically integrated the Rayleigh-Sommerfeld diffraction integral, Eq.~(\ref{rayleigh}), for structures with finite extent. This allows to study the effects due to the finite radial size of the spiral. On the other hand, numerical simulations using a 3D pseudo-spectral time-domain method using a $k$-space corrector operator \cite{firouzi2012} was also performed. In these simulations, a steel spiral grating embedded in water is considered, allowing the acoustic waves to penetrate in the grating's bulk material, so the effect on the compressibility of the material was considered.

\begin{figure}[tbp]
	\centering
	\includegraphics[width=9cm]{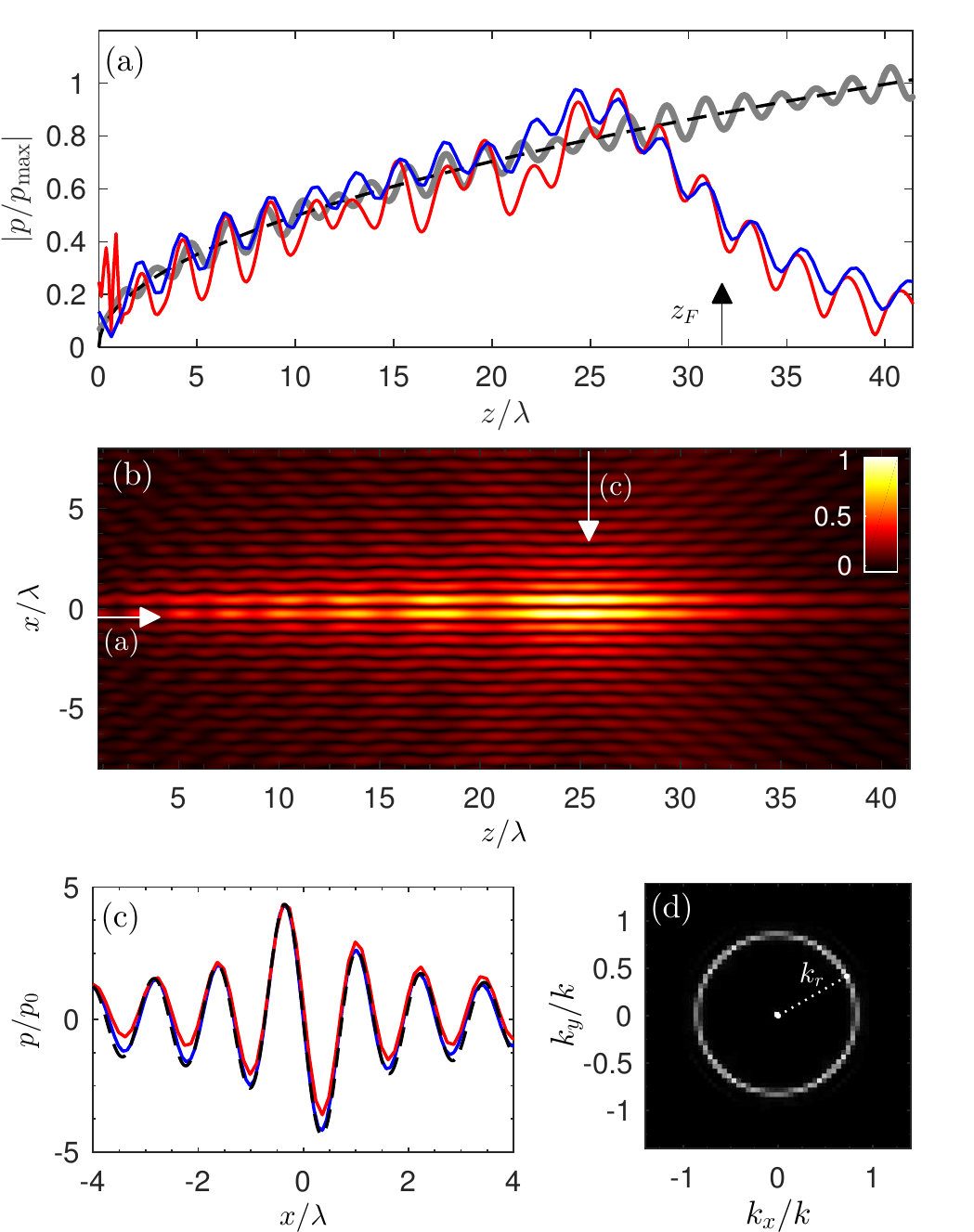}
	\caption{(a) Longitudinal pressure distribution along the first lobe of the Bessel beam obtained from Eq.~(\ref{rayleigh}) for (blue) $M=40$ and (gray) $M=1000$, (red) $k$-space simulation considering elastic scatters and (black dashed) analytic Eq.~(\ref{pressio}). (b) Pressure map distribution obtained by numerical evaluation of the Rayleigh-Sommerfeld diffraction integral, Eq.~(\ref{rayleigh}), for a spiral of $M=40$. (c) Transversal pressure distribution at $z/\lambda=25$ obtained from Eq.~(\ref{rayleigh}) for (blue) $M=40$, (red) $k$-space simulation considering elastic scatters and (black dashed) analytic Eq.~(\ref{pressio}). (d) Far-field showing the characteristic ring of the Bessel beam, where $k_{r}=2\pi /a$.}
	\label{fig:all4x}
\end{figure}

Figure \ref{fig:all4x}~(b) shows the pressure field from the spiral grating analyzed in the previous Section as obtained now numerical integration of the Rayleigh-Sommerfeld diffraction integral, Eq.~(\ref{rayleigh}), for a finite structure of $M=40$ loops of spiral ($R=40a$). It can be observed the formation of a first order Bessel beam with the elongated zero-line at the axis. An axial cross section along the first lobe of the Bessel beam is shown in Fig.~\ref{fig:all4x}~(a) (blue line). This numerical integration of the Rayleigh-Sommerfeld integral agrees well with theory (black dashed line). It is worth noting here that the longitudinal field oscillations, observed in Bessel beams generated by axicons, are not present in Eq.~(\ref{pressio}) as long this result was derived for a non-truncated spiral, $M\to \infty$. As the number spiral loops increases, the longitudinal oscillations tend to disappear and the field converges to one given by the Eq.~(\ref{pressio}). To prove this we have evaluated a spiral grating with $M'=1000$, with the same radius as the previous spiral (i.e. the distance between the scatterers, $a'$ scaled respectively $R=M'a'=40a$); gray curve in Fig. \ref{fig:all4x}~(a) clearly shows this convergence. 

Furthermore, the beam amplitude follows Eq.~(\ref{pressio}) incrasingly along $z$ from $z=0$. However, as shown in Fig. \ref{fig:all4x}(a) for the case of truncated systems (blue line) the amplitude grows up to a given distance, $z=z_F$. This distance can be geometrically estimated through the zero-th order Bessel beams as \cite{jimenez2014}
\begin{equation}
z_F=\frac{Ra}{N\lambda}\sqrt{1 - \left(\frac{N\lambda}{a}\right)^2} \,,
\end{equation}

\noindent where $N\in\mathbb{N}$ is the diffraction order and $R=Ma$ is the radius of the spiral with $M$ the windings. The $z_F$ for the spiral with $M=40$ is shown in Fig. \ref{fig:all4x}(a) while for the case with $M=1000$ is out of the limits of the plotted frequencies. The estimation of $z_F$ is in good agreement with the numerical integration of the Eq.~(\ref{rayleigh}).

In the radial direction, the beam follows a Bessel profile, Eq. (\ref{pressio}). A transversal cross section at $z/\lambda=25$, and $y=0$ is presented in Fig.~\ref{fig:all4x}~(c), showing a good agreement between the radial field distribution obtained by the numerical integration and theory. In order to prove this, Fig.~\ref{fig:all4x}~(d) shows the far-field of the truncated grating, where the characteristic ring of a Bessel beam is observed, corresponding to the wavevector $k_r=2\pi / a$.

Finally, we analyze the effect of the rigidity of the material of the spiral grating on the scattered field using numerical simulations. We analyze the scattering by the fixed spiral grating with $M=40a$ radiated by a plane wave. In these simulations, no losses were included and the thickness of the grating was $\lambda/4$. The simulations results, red curve in Fig. \ref{fig:all4x}~(a), present axial oscillations due to the finite aperture of the grating as already predicted by the direct integral of the Eq. (\ref{rayleigh}). Here, excellent agreement is found between simulations and theory. The small discrepancies are caused by the fact that the scatterers are of finite size, i. e. not the perfect punctual sources (as assumed in the Rayleigh-Sommerfeld integral), and also due to the bulk resonances into the body of the steel grating. The radial profile, shown in Fig. \ref{fig:all4x}(c), also shows the Bessel profile in good agreement with the theory and the direct integration of the Eq. (\ref{rayleigh}). Therefore, the effect of the rigidity of the material of the structure is very small.

\begin{figure*}[ht]
	\centering
	\includegraphics[width=18cm]{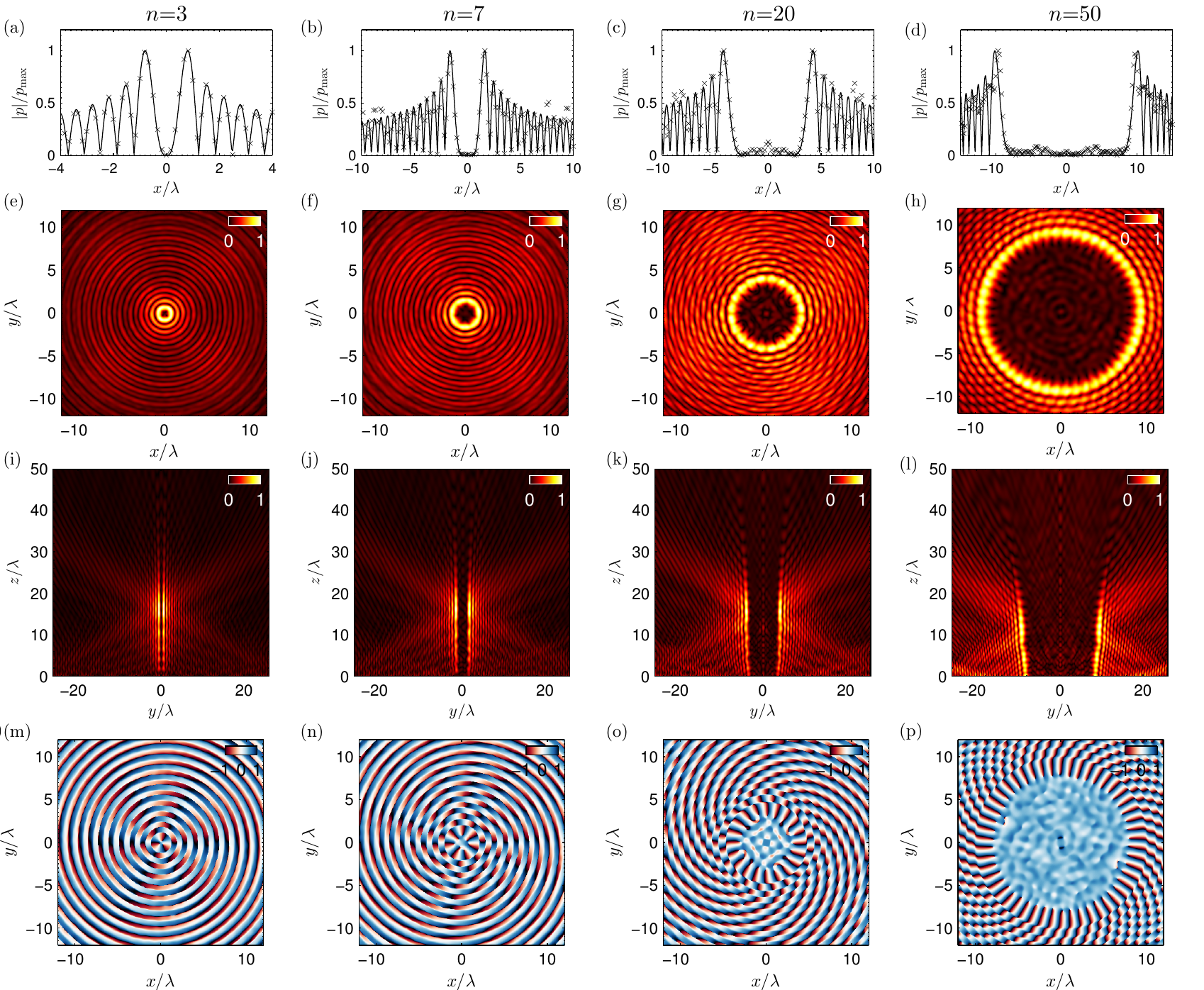}
	\caption {$n$-th order Bessel beams formed by spiral gratings of $n$ arms. (a-d) Transversal pressure distribution at $z=z_F$ and $y=0$, (continuous line) Eq.~(\ref{pressio}), (markers) $k$-space numerical solution of the wave equation assuming a steel grating embedded in water. Pressure magnitude obtained at (e-h) $z=F$ and (i-l) $x=0$. Colorbars in normalized units $p/p_{\mathrm{max}}$. (m-p) Phase of the field is calculated at $z=z_F$, colorbars in normalized units $\phi /\pi$.}
	\label{fig:arms}
\end{figure*}

\section{High order Bessel beams}
\label{sec:HOBB} 
One of the consequences of the procedure described above is that it is possible to generate HOBBs by using multiple-arm spirals ($n>$1). The parametric Eq.~(\ref{spiral1}) describes $n$ arms, separated by a fixed distance $a$. In the case of $n>1$, the rate of growth of each individual arm is increased by a factor of $n$, as it is underlined by the dark arms in Fig.~\ref{fig:Figarms}. The variation of the phase ($\phi$) with the angle ($\theta$) is therefore increased, and the conical wavefront formed by the axisymmetric grating presents a total phase shift of $2\pi n$ over a complete turn, as follows from Eq.~(\ref{velocity1}). When converging to the axis, the conical wavefront forms a Bessel beam in the same way as in the previous Section. However, as the phase rotation is increased, the vortex presents a topological charge of $n$. The conformed Bessel beam is therefore a $n$-th order Bessel beam and as a consequence the hollow central area of the beam is extended.

\begin{figure*}[ht]
	\centering
	\includegraphics[width=18cm]{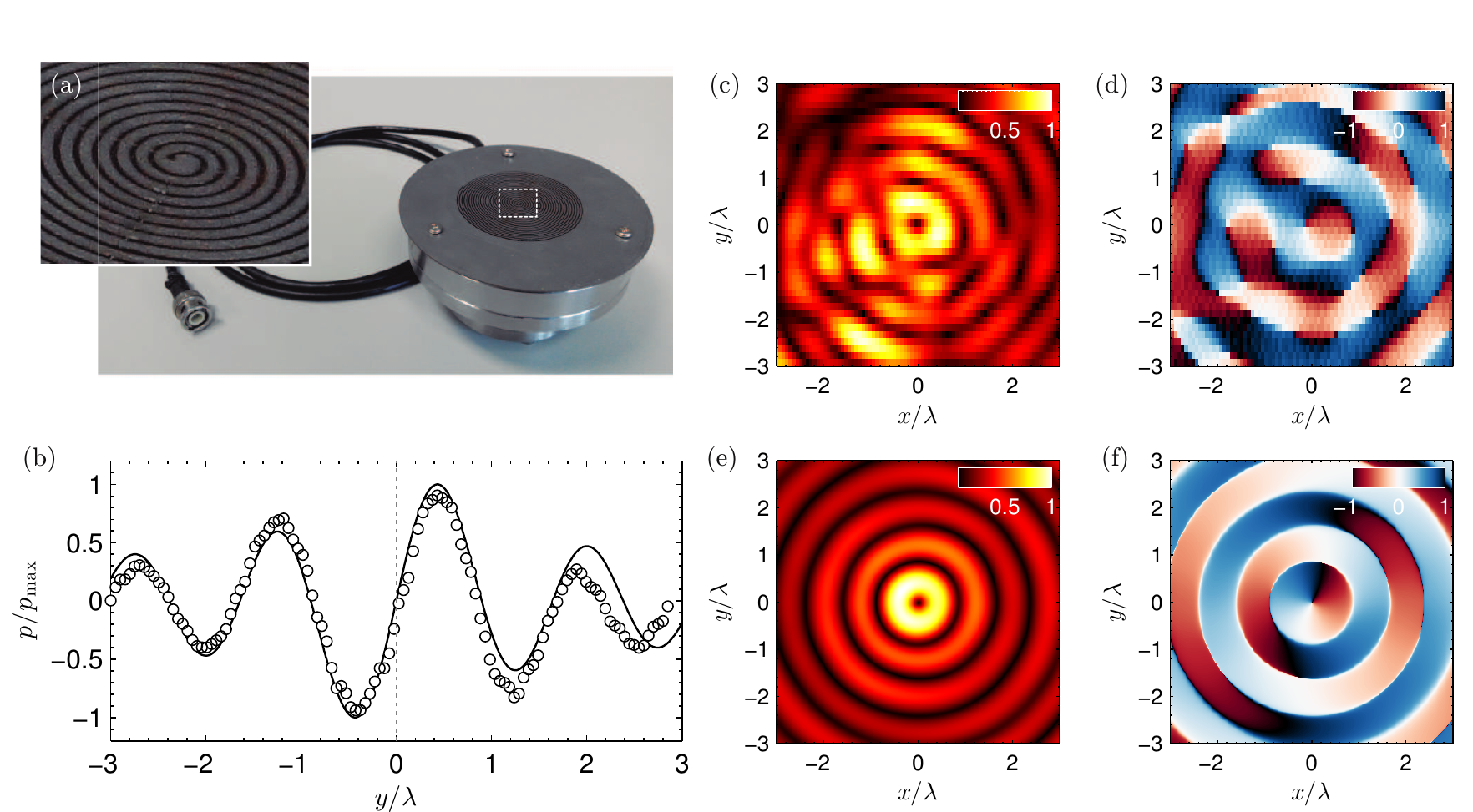}
	\caption{(a) Spiral grating used in experiments. Inset shows a zoom were the spiral can be clearly seen. (b) Transversal cross-section of the pressure field at a distance $z=10/\lambda$ normalized to the maximum for (solid line) the analytic expression of the first-order Bessel beam and (circles) the experimental results. The vertical dotted line represents the position where the  field is null. (c-d) Numerical predictions and (e-f) experimental results of the magnitude and phase, respectively. Colorbars in normalized units ($p/p_0$ for the field magnitude and $\phi/\pi$ for the phase).}
	\label{fig:expA}
\end{figure*}

Figure \ref{fig:arms} presents the formation of HOBBs for the cases $n=$ 3, 7, 20 and 50. In the Figs.~\ref{fig:arms}(a-d), the transversal pressure distribution at $z=z_F$ and $y=0$ is presented. In this case, the solid line presents the analytical solution for the transverse field of an ideal Bessel beam, $J_n(k_r r)$, and symbols the $k$-space numerical solution assuming a steel grating embedded in water. The simulations are in agreement with theory, even in the case of $n=50$. Notice here the discrepancies at high radial values, in which the approximated theoretical solution fails because corrections of higher order should be taken into account, and also the finite size effect of the sample is noticed in this regions.

The complete transversal map at $z=z_F$ is shown in Figs.~\ref{fig:arms}~(e-h). The areas of strongly reduced sound around the symmetry axis appear the larger is "n", the larger are the zero-field areas.
The radius of the reduced sound areas can be estimated from the position of the first maximum of the $n$-th Bessel function as
\begin{eqnarray}
r_{n}=\frac{j'_{n}a}{2\pi}\simeq\frac{(n+0.8086n^{1/3})a}{2\pi},
\end{eqnarray}
where $j'_{n}$ is the first zero of the first derivative of the $n$-th Bessel function \cite{Abramowich}. For the case of $n=7$, $j'_n\simeq8.57$, therefore $r_7/\lambda=1.64$ in agreement with the results shown in Fig. \ref{fig:arms}(b). The axial map of the field shown in Fig.~\ref{fig:arms}~(i-l), also shows the dependence of the hollow central part of the beam on the increasing order $n$. Of special interest is the generation of high order beams, e.g. $n=50$. In this case, $j'_n\simeq57.12$, therefore the zero in the center of the hollow beam covers a cylindrical volume with a diameter of $2r_{50}/\lambda=21.82$, in the interior of which the scattered sound is almost absent. 

Finally, the phase of the field is presented in the sub-panels Fig.~\ref{fig:arms}~(m-p) at $z=z_F$. It can be seen that the number of times the phase rotates in each turn, i.e. the topological charge, is proportional to the order of the Bessel beam, in accordance with Eq.~(\ref{pressio}). The formed field are therefore vortex beams of topological charge $n$, where the topological charge of the vortex can be controlled directly by the number of arms of the spiral grating.

These results show that HOBBs can be generated even by acoustically permeable gratings for the water/steel impedance contrast ratio, i.e. under realistic conditions for common ultrasound applications. 

\section{Experimental validation: A first order Bessel beam formed by a spiral grating}\label{sec:exp}

A spiral grating in water is experimentally studied in order to create an acoustic vortex by this kind of passive elements. A spiral profile in a stainless steel plate of 0.8 mm thickness was laser cutted. The diameter of the scattering area was $\Delta r=0.75$ mm and the grating period $a=1$ mm. The width of the open slits is $a-\Delta r=0.25$ mm of the open slits. The spiral winds $M=20$ times and total radius is $R=50$ mm. The spiral plate is aligned and placed in front of a flat ultrasonic transducer of the same diameter as the grating, as can be seen in Fig.~\ref{fig:expA}~(a). The source was driven by a 50 cycles sinusoidal pulse burst of frequency $f_0=2.22$ MHz using a function generator (14 bits, 100 MS/s, model PXI5412, National Instruments) and a linear RF ampliffer (ENI 1040L, 400W, 55dB, ENI, Rochester, NY). The pressure waveforms were recorded with the help of a HNR 500 $\mathrm{\mu}$m needle PVDF hydrophone (Onda Corp, CA), and a digitizer (64 MS/s, model PXI5620, National Instruments) was used. A three-axis micropositioning system (OWIS GmbH, Germany) was used to move the hydrophone in three orthogonal directions with an accuracy of 10 $\mathrm{\mu}$m and a National Instruments PXI-Technology controller NI8176 was used to control all the devices. The distance between the grating and the source plane was adjusted to 0.5 mm. 

The transversal cross section of the pressure field is shown in Fig.~\ref{fig:expA}~(b). The transversal cross-section in the experiments (circles) was chosen for an azimuthal angle in the $(y, x)$ plane at $\theta=-3^\circ$. The traversal profile agrees well with the shape of the first order Bessel beam (continuous black line). Although minor differences are visible for the lobe amplitude on the positive axis $y$, the main features of the HOBB, i.e. its central zero and rotational vortex, are correctly reproduced by the proposed experimental setup.

Figures \ref{fig:expA}~(c-f) show the experimental measurements and numerical predictions of the acoustic field in a transverse plane to the axis at $z=z_F$. Figure \ref{fig:expA}~(c) shows the experimental measurement for the amplitude of the field formed by the spiral grating. The pattern matches the characteristic first order Bessel beam with null amplitude in axis and a set of rings of pressure maxima with increasing radii. Figure \ref{fig:expA}~(e) presents the corresponding numerical predictions. Both results show good agreement. Some differences appear, mainly due to technical imperfections: the misalignment between the grating and the source plane, the nonuniform vibration of the piezoelectric transducer and the resonances between the source and the grating. The phase of the field is presented in Figs.~\ref{fig:expA}~(d) and \ref{fig:expA}~(f) for experimental results and numerical predictions respectively. The characteristic screw phase dislocation at the center was observed: a complete loop around a point centered on the axis represents a linear and continuous variation of the phase from $0$ to $2\pi$, i.e. the topological charge of the acoustic vortex is one. Remark that a shift of $\pi$ in phase is observed between any point and its image with respect to the central axis. This a proof that the wave transmitted through the grating is therefore an acoustic vortex. 

\section{Conclusions}

The formation of HOBBs by scattering of plane waves on an Archimedes' spiral grating is theoretically and experimentally reported in this work . The effect of the finite size of the sample is analyzed numerically. The main result is that, due to finite size of the sample, the HOBBs are truncated. All the beams analyzed are characterized by a zero of the pressure field along the $z$-axis, i. e. the vortex line. The size of this hollow part of the beam is dependent on the topological charge of the HOBB, which is directly controlled by the number of arms of the spiral. Experimental tests in the ultrasound regime have been performed showing the case of a truncated-first order Bessel beam. Good agreement between theory, experimental measurements and numerical simulations is found for the acoustic pressure field amplitude, as well as for the screw phase dislocations. Therefore, the system shown in this work seems to be of special interest for the generation of arbitrary $n$th-order Bessel beams using regular spiral patters with $n$-arms. 

The system shown in this work to synthesize HOBBs presents a high potential in ultrasound particle manipulation techniques and, in general, in acoustic radiation force applications in which the HOBBs have attracted great interest. This method provides the possibility of generation of Bessel beams of arbitrary order by a passive and cheap device if compared with acoustical vortices generated by active arrays of transducers. The generation of an acoustic vortex by arrays is limited by the amount of active elements and its size. In contrast, the beam resolution by the proposed setting is limited only by the ratio between the wavelength and the spacing between slits. Nowadays, with the increase in the performance of 3D printing and laser cutting techniques, the conformation of HOBBs by spiral gratings offer an alternative to multielement transducers to generate acoustical vortices.

\begin{acknowledgments} 
The work was supported by ESA Contract 4000110685/14/NL/SC. LMGR Acknowledges support by MINECO and FEDER, under Grant MTM2012-36740-c02-02. NJ Acknowledges finncial support from PAID-2011 Universitat Polit\`ecnica de Val\`encia.
\end{acknowledgments}


\begin{thebibliography}{40}%
\makeatletter
\providecommand \@ifxundefined [1]{%
 \@ifx{#1\undefined}
}%
\providecommand \@ifnum [1]{%
 \ifnum #1\expandafter \@firstoftwo
 \else \expandafter \@secondoftwo
 \fi
}%
\providecommand \@ifx [1]{%
 \ifx #1\expandafter \@firstoftwo
 \else \expandafter \@secondoftwo
 \fi
}%
\providecommand \natexlab [1]{#1}%
\providecommand \enquote  [1]{``#1''}%
\providecommand \bibnamefont  [1]{#1}%
\providecommand \bibfnamefont [1]{#1}%
\providecommand \citenamefont [1]{#1}%
\providecommand \href@noop [0]{\@secondoftwo}%
\providecommand \href [0]{\begingroup \@sanitize@url \@href}%
\providecommand \@href[1]{\@@startlink{#1}\@@href}%
\providecommand \@@href[1]{\endgroup#1\@@endlink}%
\providecommand \@sanitize@url [0]{\catcode `\\12\catcode `\$12\catcode
  `\&12\catcode `\#12\catcode `\^12\catcode `\_12\catcode `\%12\relax}%
\providecommand \@@startlink[1]{}%
\providecommand \@@endlink[0]{}%
\providecommand \url  [0]{\begingroup\@sanitize@url \@url }%
\providecommand \@url [1]{\endgroup\@href {#1}{\urlprefix }}%
\providecommand \urlprefix  [0]{URL }%
\providecommand \Eprint [0]{\href }%
\providecommand \doibase [0]{http://dx.doi.org/}%
\providecommand \selectlanguage [0]{\@gobble}%
\providecommand \bibinfo  [0]{\@secondoftwo}%
\providecommand \bibfield  [0]{\@secondoftwo}%
\providecommand \translation [1]{[#1]}%
\providecommand \BibitemOpen [0]{}%
\providecommand \bibitemStop [0]{}%
\providecommand \bibitemNoStop [0]{.\EOS\space}%
\providecommand \EOS [0]{\spacefactor3000\relax}%
\providecommand \BibitemShut  [1]{\csname bibitem#1\endcsname}%
\let\auto@bib@innerbib\@empty
\bibitem [{\citenamefont {Durnin}\ \emph {et~al.}(1987)\citenamefont {Durnin},
  \citenamefont {Miceli~Jr},\ and\ \citenamefont {Eberly}}]{durnin1987}%
  \BibitemOpen
  \bibfield  {author} {\bibinfo {author} {\bibfnamefont {J.}~\bibnamefont
  {Durnin}}, \bibinfo {author} {\bibfnamefont {J.}~\bibnamefont {Miceli~Jr}}, \
  and\ \bibinfo {author} {\bibfnamefont {J.}~\bibnamefont {Eberly}},\
  }\href@noop {} {\bibfield  {journal} {\bibinfo  {journal} {Phys. Rev. Lett.}\
  }\textbf {\bibinfo {volume} {58}},\ \bibinfo {pages} {1499} (\bibinfo {year}
  {1987})}\BibitemShut {NoStop}%
\bibitem [{\citenamefont {Vasara}\ \emph {et~al.}(1989)\citenamefont {Vasara},
  \citenamefont {Turunen},\ and\ \citenamefont {Friberg}}]{vasara1989}%
  \BibitemOpen
  \bibfield  {author} {\bibinfo {author} {\bibfnamefont {A.}~\bibnamefont
  {Vasara}}, \bibinfo {author} {\bibfnamefont {J.}~\bibnamefont {Turunen}}, \
  and\ \bibinfo {author} {\bibfnamefont {A.~T.}\ \bibnamefont {Friberg}},\
  }\href@noop {} {\bibfield  {journal} {\bibinfo  {journal} {J. Opt. Soc. Am.
  A}\ }\textbf {\bibinfo {volume} {6}},\ \bibinfo {pages} {1748} (\bibinfo
  {year} {1989})}\BibitemShut {NoStop}%
\bibitem [{\citenamefont {Arlt}\ and\ \citenamefont
  {Dholakia}(2000)}]{arlt2000}%
  \BibitemOpen
  \bibfield  {author} {\bibinfo {author} {\bibfnamefont {J.}~\bibnamefont
  {Arlt}}\ and\ \bibinfo {author} {\bibfnamefont {K.}~\bibnamefont
  {Dholakia}},\ }\href@noop {} {\bibfield  {journal} {\bibinfo  {journal} {Opt.
  Commun.}\ }\textbf {\bibinfo {volume} {177}},\ \bibinfo {pages} {297}
  (\bibinfo {year} {2000})}\BibitemShut {NoStop}%
\bibitem [{\citenamefont {Oemrawsingh}\ \emph {et~al.}(2004)\citenamefont
  {Oemrawsingh}, \citenamefont {van Houwelingen}, \citenamefont {Eliel},
  \citenamefont {Woerdman}, \citenamefont {Verstegen}, \citenamefont
  {Kloosterboer},\ and\ \citenamefont {'t~Hooft}}]{Oemrawsingh2004}%
  \BibitemOpen
  \bibfield  {author} {\bibinfo {author} {\bibfnamefont {S.~S.~R.}\
  \bibnamefont {Oemrawsingh}}, \bibinfo {author} {\bibfnamefont {J.~A.~W.}\
  \bibnamefont {van Houwelingen}}, \bibinfo {author} {\bibfnamefont {E.~R.}\
  \bibnamefont {Eliel}}, \bibinfo {author} {\bibfnamefont {J.~P.}\ \bibnamefont
  {Woerdman}}, \bibinfo {author} {\bibfnamefont {E.~J.~K.}\ \bibnamefont
  {Verstegen}}, \bibinfo {author} {\bibfnamefont {J.~G.}\ \bibnamefont
  {Kloosterboer}}, \ and\ \bibinfo {author} {\bibfnamefont {G.~W.}\
  \bibnamefont {'t~Hooft}},\ }\href@noop {} {\bibfield  {journal} {\bibinfo
  {journal} {Appl. Opt.}\ }\textbf {\bibinfo {volume} {43}},\ \bibinfo {pages}
  {688} (\bibinfo {year} {2004})}\BibitemShut {NoStop}%
\bibitem [{\citenamefont {Marston}(2006)}]{marston2006}%
  \BibitemOpen
  \bibfield  {author} {\bibinfo {author} {\bibfnamefont {P.~L.}\ \bibnamefont
  {Marston}},\ }\href@noop {} {\bibfield  {journal} {\bibinfo  {journal} {J.
  Acoust. Soc. Am.}\ }\textbf {\bibinfo {volume} {120}},\ \bibinfo {pages}
  {3518} (\bibinfo {year} {2006})}\BibitemShut {NoStop}%
\bibitem [{\citenamefont {Mitri}(2009{\natexlab{a}})}]{mitri2009}%
  \BibitemOpen
  \bibfield  {author} {\bibinfo {author} {\bibfnamefont {F.}~\bibnamefont
  {Mitri}},\ }\href@noop {} {\bibfield  {journal} {\bibinfo  {journal} {J.
  Phys. A: Math. Theor.}\ }\textbf {\bibinfo {volume} {42}},\ \bibinfo {pages}
  {245202} (\bibinfo {year} {2009}{\natexlab{a}})}\BibitemShut {NoStop}%
\bibitem [{\citenamefont {Mitri}(2009{\natexlab{b}})}]{mitri2009b}%
  \BibitemOpen
  \bibfield  {author} {\bibinfo {author} {\bibfnamefont {F.}~\bibnamefont
  {Mitri}},\ }\href@noop {} {\bibfield  {journal} {\bibinfo  {journal}
  {Ultrasonics}\ }\textbf {\bibinfo {volume} {49}},\ \bibinfo {pages} {794}
  (\bibinfo {year} {2009}{\natexlab{b}})}\BibitemShut {NoStop}%
\bibitem [{\citenamefont {Jim{\'e}nez}\ \emph {et~al.}(2014)\citenamefont
  {Jim{\'e}nez}, \citenamefont {Romero-Garc{\'\i}a}, \citenamefont {Pic{\'o}},
  \citenamefont {Cebrecos}, \citenamefont {S{\'a}nchez-Morcillo}, \citenamefont
  {Garcia-Raffi}, \citenamefont {S{\'a}nchez-P{\'e}rez},\ and\ \citenamefont
  {Staliunas}}]{jimenez2014}%
  \BibitemOpen
  \bibfield  {author} {\bibinfo {author} {\bibfnamefont {N.}~\bibnamefont
  {Jim{\'e}nez}}, \bibinfo {author} {\bibfnamefont {V.}~\bibnamefont
  {Romero-Garc{\'\i}a}}, \bibinfo {author} {\bibfnamefont {R.}~\bibnamefont
  {Pic{\'o}}}, \bibinfo {author} {\bibfnamefont {A.}~\bibnamefont {Cebrecos}},
  \bibinfo {author} {\bibfnamefont {V.~J.}\ \bibnamefont
  {S{\'a}nchez-Morcillo}}, \bibinfo {author} {\bibfnamefont {L.}~\bibnamefont
  {Garcia-Raffi}}, \bibinfo {author} {\bibfnamefont {J.~V.}\ \bibnamefont
  {S{\'a}nchez-P{\'e}rez}}, \ and\ \bibinfo {author} {\bibfnamefont
  {K.}~\bibnamefont {Staliunas}},\ }\href@noop {} {\bibfield  {journal}
  {\bibinfo  {journal} {Europhys. Lett.}\ }\textbf {\bibinfo {volume} {106}},\
  \bibinfo {pages} {24005} (\bibinfo {year} {2014})}\BibitemShut {NoStop}%
\bibitem [{\citenamefont {Jim{\'e}nez}\ \emph {et~al.}(2015)\citenamefont
  {Jim{\'e}nez}, \citenamefont {Romero-Garc{\'\i}a}, \citenamefont {Pic{\'o}},
  \citenamefont {Garcia-Raffi},\ and\ \citenamefont {Staliunas}}]{jimenez2015}%
  \BibitemOpen
  \bibfield  {author} {\bibinfo {author} {\bibfnamefont {N.}~\bibnamefont
  {Jim{\'e}nez}}, \bibinfo {author} {\bibfnamefont {V.}~\bibnamefont
  {Romero-Garc{\'\i}a}}, \bibinfo {author} {\bibfnamefont {R.}~\bibnamefont
  {Pic{\'o}}}, \bibinfo {author} {\bibfnamefont {L.}~\bibnamefont
  {Garcia-Raffi}}, \ and\ \bibinfo {author} {\bibfnamefont {K.}~\bibnamefont
  {Staliunas}},\ }\href@noop {} {\bibfield  {journal} {\bibinfo  {journal}
  {Appl. Phys. Lett.}\ }\textbf {\bibinfo {volume} {107}},\ \bibinfo {pages}
  {204103} (\bibinfo {year} {2015})}\BibitemShut {NoStop}%
\bibitem [{\citenamefont {Pfeiffer}\ and\ \citenamefont
  {Grbic}(2015)}]{pfeiffer2015}%
  \BibitemOpen
  \bibfield  {author} {\bibinfo {author} {\bibfnamefont {C.}~\bibnamefont
  {Pfeiffer}}\ and\ \bibinfo {author} {\bibfnamefont {A.}~\bibnamefont
  {Grbic}},\ }\href@noop {} {\bibfield  {journal} {\bibinfo  {journal} {Phys.
  Rev. B}\ }\textbf {\bibinfo {volume} {91}} (\bibinfo {year}
  {2015})}\BibitemShut {NoStop}%
\bibitem [{\citenamefont {Baresch}\ \emph {et~al.}(2016)\citenamefont
  {Baresch}, \citenamefont {Thomas},\ and\ \citenamefont
  {Marchiano}}]{Baresch16}%
  \BibitemOpen
  \bibfield  {author} {\bibinfo {author} {\bibfnamefont {D.}~\bibnamefont
  {Baresch}}, \bibinfo {author} {\bibfnamefont {J.-L.}\ \bibnamefont {Thomas}},
  \ and\ \bibinfo {author} {\bibfnamefont {R.}~\bibnamefont {Marchiano}},\
  }\href@noop {} {\bibfield  {journal} {\bibinfo  {journal} {Phys. Rev. Lett.}\
  }\textbf {\bibinfo {volume} {116}} (\bibinfo {year} {2016})}\BibitemShut
  {NoStop}%
\bibitem [{\citenamefont {Heckenberg}\ \emph {et~al.}(1992)\citenamefont
  {Heckenberg}, \citenamefont {McDuff}, \citenamefont {Smith},\ and\
  \citenamefont {White}}]{heckenberg1992}%
  \BibitemOpen
  \bibfield  {author} {\bibinfo {author} {\bibfnamefont {N.}~\bibnamefont
  {Heckenberg}}, \bibinfo {author} {\bibfnamefont {R.}~\bibnamefont {McDuff}},
  \bibinfo {author} {\bibfnamefont {C.}~\bibnamefont {Smith}}, \ and\ \bibinfo
  {author} {\bibfnamefont {A.}~\bibnamefont {White}},\ }\href@noop {}
  {\bibfield  {journal} {\bibinfo  {journal} {Opt. Lett.}\ }\textbf {\bibinfo
  {volume} {17}},\ \bibinfo {pages} {221} (\bibinfo {year} {1992})}\BibitemShut
  {NoStop}%
\bibitem [{\citenamefont {Bekshaev}\ and\ \citenamefont
  {Karamoch}(2008)}]{bekshaev2008}%
  \BibitemOpen
  \bibfield  {author} {\bibinfo {author} {\bibfnamefont {A.~Y.}\ \bibnamefont
  {Bekshaev}}\ and\ \bibinfo {author} {\bibfnamefont {A.}~\bibnamefont
  {Karamoch}},\ }\href@noop {} {\bibfield  {journal} {\bibinfo  {journal} {Opt.
  Commun.}\ }\textbf {\bibinfo {volume} {281}},\ \bibinfo {pages} {3597}
  (\bibinfo {year} {2008})}\BibitemShut {NoStop}%
\bibitem [{\citenamefont {Ashkin}(1970)}]{ashkin1970}%
  \BibitemOpen
  \bibfield  {author} {\bibinfo {author} {\bibfnamefont {A.}~\bibnamefont
  {Ashkin}},\ }\href@noop {} {\bibfield  {journal} {\bibinfo  {journal} {Phys.
  Rev. Lett.}\ }\textbf {\bibinfo {volume} {24}},\ \bibinfo {pages} {156}
  (\bibinfo {year} {1970})}\BibitemShut {NoStop}%
\bibitem [{\citenamefont {Ashkin}\ \emph {et~al.}(1987)\citenamefont {Ashkin},
  \citenamefont {Dziedzic},\ and\ \citenamefont {Yamane}}]{ashkin1987}%
  \BibitemOpen
  \bibfield  {author} {\bibinfo {author} {\bibfnamefont {A.}~\bibnamefont
  {Ashkin}}, \bibinfo {author} {\bibfnamefont {J.}~\bibnamefont {Dziedzic}}, \
  and\ \bibinfo {author} {\bibfnamefont {T.}~\bibnamefont {Yamane}},\
  }\href@noop {} {\bibfield  {journal} {\bibinfo  {journal} {Nature}\ }\textbf
  {\bibinfo {volume} {330}},\ \bibinfo {pages} {769} (\bibinfo {year}
  {1987})}\BibitemShut {NoStop}%
\bibitem [{\citenamefont {Ashkin}\ \emph {et~al.}(1986)\citenamefont {Ashkin},
  \citenamefont {Dziedzic}, \citenamefont {Bjorkholm},\ and\ \citenamefont
  {Chu}}]{ashkin1986}%
  \BibitemOpen
  \bibfield  {author} {\bibinfo {author} {\bibfnamefont {A.}~\bibnamefont
  {Ashkin}}, \bibinfo {author} {\bibfnamefont {J.}~\bibnamefont {Dziedzic}},
  \bibinfo {author} {\bibfnamefont {J.}~\bibnamefont {Bjorkholm}}, \ and\
  \bibinfo {author} {\bibfnamefont {S.}~\bibnamefont {Chu}},\ }\href@noop {}
  {\bibfield  {journal} {\bibinfo  {journal} {Opt. Lett.}\ }\textbf {\bibinfo
  {volume} {11}},\ \bibinfo {pages} {288} (\bibinfo {year} {1986})}\BibitemShut
  {NoStop}%
\bibitem [{\citenamefont {Grier}(2003)}]{grier2003}%
  \BibitemOpen
  \bibfield  {author} {\bibinfo {author} {\bibfnamefont {D.~G.}\ \bibnamefont
  {Grier}},\ }\href@noop {} {\bibfield  {journal} {\bibinfo  {journal}
  {Nature}\ }\textbf {\bibinfo {volume} {424}},\ \bibinfo {pages} {810}
  (\bibinfo {year} {2003})}\BibitemShut {NoStop}%
\bibitem [{\citenamefont {Nye}\ and\ \citenamefont {Berry}(1974)}]{nye1974}%
  \BibitemOpen
  \bibfield  {author} {\bibinfo {author} {\bibfnamefont {J.}~\bibnamefont
  {Nye}}\ and\ \bibinfo {author} {\bibfnamefont {M.}~\bibnamefont {Berry}},\
  }\bibfield  {booktitle} {\emph {\bibinfo {booktitle} {Proc. R. Soc. A}},\
  }\href@noop {} {\bibfield  {journal} {\bibinfo  {journal} {Proc. R. Soc.
  London, Ser. A}\ }\textbf {\bibinfo {volume} {336}},\ \bibinfo {pages} {165}
  (\bibinfo {year} {1974})}\BibitemShut {NoStop}%
\bibitem [{\citenamefont {Hefner}\ and\ \citenamefont
  {Marston}(1999)}]{hefner1999}%
  \BibitemOpen
  \bibfield  {author} {\bibinfo {author} {\bibfnamefont {B.~T.}\ \bibnamefont
  {Hefner}}\ and\ \bibinfo {author} {\bibfnamefont {P.~L.}\ \bibnamefont
  {Marston}},\ }\href@noop {} {\bibfield  {journal} {\bibinfo  {journal} {J.
  Acoust. Soc. Am.}\ }\textbf {\bibinfo {volume} {106}},\ \bibinfo {pages}
  {3313} (\bibinfo {year} {1999})}\BibitemShut {NoStop}%
\bibitem [{\citenamefont {Gspan}\ \emph {et~al.}(2004)\citenamefont {Gspan},
  \citenamefont {Meyer}, \citenamefont {Bernet},\ and\ \citenamefont
  {Ritsch-Marte}}]{gspan2004}%
  \BibitemOpen
  \bibfield  {author} {\bibinfo {author} {\bibfnamefont {S.}~\bibnamefont
  {Gspan}}, \bibinfo {author} {\bibfnamefont {A.}~\bibnamefont {Meyer}},
  \bibinfo {author} {\bibfnamefont {S.}~\bibnamefont {Bernet}}, \ and\ \bibinfo
  {author} {\bibfnamefont {M.}~\bibnamefont {Ritsch-Marte}},\ }\href@noop {}
  {\bibfield  {journal} {\bibinfo  {journal} {J. Acoust. Soc. Am.}\ }\textbf
  {\bibinfo {volume} {115}},\ \bibinfo {pages} {1142} (\bibinfo {year}
  {2004})}\BibitemShut {NoStop}%
\bibitem [{\citenamefont {Thomas}\ and\ \citenamefont
  {Marchiano}(2003)}]{thomas2003}%
  \BibitemOpen
  \bibfield  {author} {\bibinfo {author} {\bibfnamefont {J.-L.}\ \bibnamefont
  {Thomas}}\ and\ \bibinfo {author} {\bibfnamefont {R.}~\bibnamefont
  {Marchiano}},\ }\href@noop {} {\bibfield  {journal} {\bibinfo  {journal}
  {Phys. Rev. Lett.}\ }\textbf {\bibinfo {volume} {91}},\ \bibinfo {pages}
  {244302} (\bibinfo {year} {2003})}\BibitemShut {NoStop}%
\bibitem [{\citenamefont {Mitri}\ \emph {et~al.}(2012)\citenamefont {Mitri},
  \citenamefont {Lobo},\ and\ \citenamefont {Silva}}]{mitri2012}%
  \BibitemOpen
  \bibfield  {author} {\bibinfo {author} {\bibfnamefont {F.}~\bibnamefont
  {Mitri}}, \bibinfo {author} {\bibfnamefont {T.}~\bibnamefont {Lobo}}, \ and\
  \bibinfo {author} {\bibfnamefont {G.}~\bibnamefont {Silva}},\ }\href@noop {}
  {\bibfield  {journal} {\bibinfo  {journal} {Phys. Rev. E}\ }\textbf {\bibinfo
  {volume} {85}},\ \bibinfo {pages} {026602} (\bibinfo {year}
  {2012})}\BibitemShut {NoStop}%
\bibitem [{\citenamefont {Hong}\ \emph {et~al.}(2015)\citenamefont {Hong},
  \citenamefont {Zhang},\ and\ \citenamefont {Drinkwater}}]{hong2015}%
  \BibitemOpen
  \bibfield  {author} {\bibinfo {author} {\bibfnamefont {Z.}~\bibnamefont
  {Hong}}, \bibinfo {author} {\bibfnamefont {J.}~\bibnamefont {Zhang}}, \ and\
  \bibinfo {author} {\bibfnamefont {B.~W.}\ \bibnamefont {Drinkwater}},\
  }\href@noop {} {\bibfield  {journal} {\bibinfo  {journal} {Phys. Rev. Lett.}\
  }\textbf {\bibinfo {volume} {114}},\ \bibinfo {pages} {214301} (\bibinfo
  {year} {2015})}\BibitemShut {NoStop}%
\bibitem [{\citenamefont {Marston}(2007)}]{marston2007}%
  \BibitemOpen
  \bibfield  {author} {\bibinfo {author} {\bibfnamefont {P.~L.}\ \bibnamefont
  {Marston}},\ }\href@noop {} {\bibfield  {journal} {\bibinfo  {journal} {J.
  Acoust. Soc. Am.}\ }\textbf {\bibinfo {volume} {122}},\ \bibinfo {pages}
  {3162} (\bibinfo {year} {2007})}\BibitemShut {NoStop}%
\bibitem [{\citenamefont {Mitri}(2008)}]{mitri2008}%
  \BibitemOpen
  \bibfield  {author} {\bibinfo {author} {\bibfnamefont {F.}~\bibnamefont
  {Mitri}},\ }\href@noop {} {\bibfield  {journal} {\bibinfo  {journal} {Ann.
  Phys.}\ }\textbf {\bibinfo {volume} {323}},\ \bibinfo {pages} {2840}
  (\bibinfo {year} {2008})}\BibitemShut {NoStop}%
\bibitem [{\citenamefont {Baresch}\ \emph {et~al.}(2013)\citenamefont
  {Baresch}, \citenamefont {Thomas},\ and\ \citenamefont
  {Marchiano}}]{baresch2013}%
  \BibitemOpen
  \bibfield  {author} {\bibinfo {author} {\bibfnamefont {D.}~\bibnamefont
  {Baresch}}, \bibinfo {author} {\bibfnamefont {J.-L.}\ \bibnamefont {Thomas}},
  \ and\ \bibinfo {author} {\bibfnamefont {R.}~\bibnamefont {Marchiano}},\
  }\href@noop {} {\bibfield  {journal} {\bibinfo  {journal} {J. Acoust. Soc.
  Am.}\ }\textbf {\bibinfo {volume} {133}},\ \bibinfo {pages} {25} (\bibinfo
  {year} {2013})}\BibitemShut {NoStop}%
\bibitem [{\citenamefont {Demore}\ \emph {et~al.}(2011)\citenamefont {Demore},
  \citenamefont {Yang}, \citenamefont {Volovick}, \citenamefont {Wang},
  \citenamefont {Cochran}, \citenamefont {MacDonald},\ and\ \citenamefont
  {Spalding}}]{demore2011}%
  \BibitemOpen
  \bibfield  {author} {\bibinfo {author} {\bibfnamefont {C.}~\bibnamefont
  {Demore}}, \bibinfo {author} {\bibfnamefont {Z.}~\bibnamefont {Yang}},
  \bibinfo {author} {\bibfnamefont {A.}~\bibnamefont {Volovick}}, \bibinfo
  {author} {\bibfnamefont {H.}~\bibnamefont {Wang}}, \bibinfo {author}
  {\bibfnamefont {S.}~\bibnamefont {Cochran}}, \bibinfo {author} {\bibfnamefont
  {M.}~\bibnamefont {MacDonald}}, \ and\ \bibinfo {author} {\bibfnamefont
  {G.}~\bibnamefont {Spalding}},\ }in\ \href@noop {} {\emph {\bibinfo
  {booktitle} {Ultrasonics Symposium (IUS), 2011 IEEE International}}}\
  (\bibinfo {organization} {IEEE},\ \bibinfo {year} {2011})\ pp.\ \bibinfo
  {pages} {180--183}\BibitemShut {NoStop}%
\bibitem [{\citenamefont {Wu}(1991)}]{wu1991}%
  \BibitemOpen
  \bibfield  {author} {\bibinfo {author} {\bibfnamefont {J.}~\bibnamefont
  {Wu}},\ }\href@noop {} {\bibfield  {journal} {\bibinfo  {journal} {J. Acoust.
  Soc. Am.}\ }\textbf {\bibinfo {volume} {89}},\ \bibinfo {pages} {2140}
  (\bibinfo {year} {1991})}\BibitemShut {NoStop}%
\bibitem [{\citenamefont {Wang}\ \emph {et~al.}(2015)\citenamefont {Wang},
  \citenamefont {Ke}, \citenamefont {Xu}, \citenamefont {Feng}, \citenamefont
  {Qiu},\ and\ \citenamefont {Liu}}]{wang2015}%
  \BibitemOpen
  \bibfield  {author} {\bibinfo {author} {\bibfnamefont {T.}~\bibnamefont
  {Wang}}, \bibinfo {author} {\bibfnamefont {M.}~\bibnamefont {Ke}}, \bibinfo
  {author} {\bibfnamefont {S.}~\bibnamefont {Xu}}, \bibinfo {author}
  {\bibfnamefont {J.}~\bibnamefont {Feng}}, \bibinfo {author} {\bibfnamefont
  {C.}~\bibnamefont {Qiu}}, \ and\ \bibinfo {author} {\bibfnamefont
  {Z.}~\bibnamefont {Liu}},\ }\href@noop {} {\bibfield  {journal} {\bibinfo
  {journal} {Appl. Phys. Lett.}\ }\textbf {\bibinfo {volume} {106}},\ \bibinfo
  {pages} {163504} (\bibinfo {year} {2015})}\BibitemShut {NoStop}%
\bibitem [{\citenamefont {Marzo}\ \emph {et~al.}(2015)\citenamefont {Marzo},
  \citenamefont {Seah}, \citenamefont {Drinkwater}, \citenamefont {Sahoo},
  \citenamefont {Long},\ and\ \citenamefont {Subramanian}}]{marzo2015}%
  \BibitemOpen
  \bibfield  {author} {\bibinfo {author} {\bibfnamefont {A.}~\bibnamefont
  {Marzo}}, \bibinfo {author} {\bibfnamefont {S.~A.}\ \bibnamefont {Seah}},
  \bibinfo {author} {\bibfnamefont {B.~W.}\ \bibnamefont {Drinkwater}},
  \bibinfo {author} {\bibfnamefont {D.~R.}\ \bibnamefont {Sahoo}}, \bibinfo
  {author} {\bibfnamefont {B.}~\bibnamefont {Long}}, \ and\ \bibinfo {author}
  {\bibfnamefont {S.}~\bibnamefont {Subramanian}},\ }\href@noop {} {\bibfield
  {journal} {\bibinfo  {journal} {Nat. Commun.}\ }\textbf {\bibinfo {volume}
  {6}} (\bibinfo {year} {2015})}\BibitemShut {NoStop}%
\bibitem [{\citenamefont {Volke-Sep{\'u}lveda}\ \emph
  {et~al.}(2008)\citenamefont {Volke-Sep{\'u}lveda}, \citenamefont
  {Santill{\'a}n},\ and\ \citenamefont {Boullosa}}]{volke2008}%
  \BibitemOpen
  \bibfield  {author} {\bibinfo {author} {\bibfnamefont {K.}~\bibnamefont
  {Volke-Sep{\'u}lveda}}, \bibinfo {author} {\bibfnamefont {A.~O.}\
  \bibnamefont {Santill{\'a}n}}, \ and\ \bibinfo {author} {\bibfnamefont
  {R.~R.}\ \bibnamefont {Boullosa}},\ }\href@noop {} {\bibfield  {journal}
  {\bibinfo  {journal} {Phys. Rev. Lett.}\ }\textbf {\bibinfo {volume} {100}},\
  \bibinfo {pages} {024302} (\bibinfo {year} {2008})}\BibitemShut {NoStop}%
\bibitem [{\citenamefont {Skeldon}\ \emph {et~al.}(2008)\citenamefont
  {Skeldon}, \citenamefont {Wilson}, \citenamefont {Edgar},\ and\ \citenamefont
  {Padgett}}]{skeldon2008}%
  \BibitemOpen
  \bibfield  {author} {\bibinfo {author} {\bibfnamefont {K.}~\bibnamefont
  {Skeldon}}, \bibinfo {author} {\bibfnamefont {C.}~\bibnamefont {Wilson}},
  \bibinfo {author} {\bibfnamefont {M.}~\bibnamefont {Edgar}}, \ and\ \bibinfo
  {author} {\bibfnamefont {M.}~\bibnamefont {Padgett}},\ }\href@noop {}
  {\bibfield  {journal} {\bibinfo  {journal} {New J. Phys.}\ }\textbf {\bibinfo
  {volume} {10}},\ \bibinfo {pages} {013018} (\bibinfo {year}
  {2008})}\BibitemShut {NoStop}%
\bibitem [{\citenamefont {Yoon}\ \emph {et~al.}(2014)\citenamefont {Yoon},
  \citenamefont {Kang}, \citenamefont {Lee}, \citenamefont {Kim},\ and\
  \citenamefont {Shung}}]{yoon2014}%
  \BibitemOpen
  \bibfield  {author} {\bibinfo {author} {\bibfnamefont {C.}~\bibnamefont
  {Yoon}}, \bibinfo {author} {\bibfnamefont {B.~J.}\ \bibnamefont {Kang}},
  \bibinfo {author} {\bibfnamefont {C.}~\bibnamefont {Lee}}, \bibinfo {author}
  {\bibfnamefont {H.~H.}\ \bibnamefont {Kim}}, \ and\ \bibinfo {author}
  {\bibfnamefont {K.~K.}\ \bibnamefont {Shung}},\ }\href@noop {} {\bibfield
  {journal} {\bibinfo  {journal} {Appl. Phys. Lett.}\ }\textbf {\bibinfo
  {volume} {105}},\ \bibinfo {pages} {214103} (\bibinfo {year}
  {2014})}\BibitemShut {NoStop}%
\bibitem [{\citenamefont {Li}\ \emph {et~al.}(2015)\citenamefont {Li},
  \citenamefont {Mao}, \citenamefont {Peng}, \citenamefont {Zhou},
  \citenamefont {Chen}, \citenamefont {Huang}, \citenamefont {Truica},
  \citenamefont {Drabick}, \citenamefont {El-Deiry}, \citenamefont {Dao} \emph
  {et~al.}}]{li2015}%
  \BibitemOpen
  \bibfield  {author} {\bibinfo {author} {\bibfnamefont {P.}~\bibnamefont
  {Li}}, \bibinfo {author} {\bibfnamefont {Z.}~\bibnamefont {Mao}}, \bibinfo
  {author} {\bibfnamefont {Z.}~\bibnamefont {Peng}}, \bibinfo {author}
  {\bibfnamefont {L.}~\bibnamefont {Zhou}}, \bibinfo {author} {\bibfnamefont
  {Y.}~\bibnamefont {Chen}}, \bibinfo {author} {\bibfnamefont {P.-H.}\
  \bibnamefont {Huang}}, \bibinfo {author} {\bibfnamefont {C.~I.}\ \bibnamefont
  {Truica}}, \bibinfo {author} {\bibfnamefont {J.~J.}\ \bibnamefont {Drabick}},
  \bibinfo {author} {\bibfnamefont {W.~S.}\ \bibnamefont {El-Deiry}}, \bibinfo
  {author} {\bibfnamefont {M.}~\bibnamefont {Dao}},  \emph {et~al.},\
  }\href@noop {} {\bibfield  {journal} {\bibinfo  {journal} {Proc. Natl. Acad.
  Sci. U.S.A.}\ }\textbf {\bibinfo {volume} {112}},\ \bibinfo {pages} {4970}
  (\bibinfo {year} {2015})}\BibitemShut {NoStop}%
\bibitem [{\citenamefont {Guo}\ \emph {et~al.}(2016)\citenamefont {Guo},
  \citenamefont {Mao}, \citenamefont {Chen}, \citenamefont {Xie}, \citenamefont
  {Lata}, \citenamefont {Li}, \citenamefont {Ren}, \citenamefont {Liu},
  \citenamefont {Yang}, \citenamefont {Dao} \emph {et~al.}}]{guo2016}%
  \BibitemOpen
  \bibfield  {author} {\bibinfo {author} {\bibfnamefont {F.}~\bibnamefont
  {Guo}}, \bibinfo {author} {\bibfnamefont {Z.}~\bibnamefont {Mao}}, \bibinfo
  {author} {\bibfnamefont {Y.}~\bibnamefont {Chen}}, \bibinfo {author}
  {\bibfnamefont {Z.}~\bibnamefont {Xie}}, \bibinfo {author} {\bibfnamefont
  {J.~P.}\ \bibnamefont {Lata}}, \bibinfo {author} {\bibfnamefont
  {P.}~\bibnamefont {Li}}, \bibinfo {author} {\bibfnamefont {L.}~\bibnamefont
  {Ren}}, \bibinfo {author} {\bibfnamefont {J.}~\bibnamefont {Liu}}, \bibinfo
  {author} {\bibfnamefont {J.}~\bibnamefont {Yang}}, \bibinfo {author}
  {\bibfnamefont {M.}~\bibnamefont {Dao}},  \emph {et~al.},\ }\href@noop {}
  {\bibfield  {journal} {\bibinfo  {journal} {Proc. Natl. Acad. Sci. U.S.A.}\
  }\textbf {\bibinfo {volume} {113}},\ \bibinfo {pages} {1522} (\bibinfo {year}
  {2016})}\BibitemShut {NoStop}%
\bibitem [{\citenamefont {Raiton}\ \emph {et~al.}(2012)\citenamefont {Raiton},
  \citenamefont {McLaughlan}, \citenamefont {Harput}, \citenamefont {Smith},
  \citenamefont {Cowell},\ and\ \citenamefont {Freear}}]{raiton2012}%
  \BibitemOpen
  \bibfield  {author} {\bibinfo {author} {\bibfnamefont {B.}~\bibnamefont
  {Raiton}}, \bibinfo {author} {\bibfnamefont {J.}~\bibnamefont {McLaughlan}},
  \bibinfo {author} {\bibfnamefont {S.}~\bibnamefont {Harput}}, \bibinfo
  {author} {\bibfnamefont {P.}~\bibnamefont {Smith}}, \bibinfo {author}
  {\bibfnamefont {D.}~\bibnamefont {Cowell}}, \ and\ \bibinfo {author}
  {\bibfnamefont {S.}~\bibnamefont {Freear}},\ }\href@noop {} {\bibfield
  {journal} {\bibinfo  {journal} {Appl. Phys. Lett.}\ }\textbf {\bibinfo
  {volume} {101}},\ \bibinfo {pages} {044102} (\bibinfo {year}
  {2012})}\BibitemShut {NoStop}%
\bibitem [{\citenamefont {Brunet}\ \emph {et~al.}(2009)\citenamefont {Brunet},
  \citenamefont {Thomas}, \citenamefont {Marchiano},\ and\ \citenamefont
  {Coulouvrat}}]{brunet2009}%
  \BibitemOpen
  \bibfield  {author} {\bibinfo {author} {\bibfnamefont {T.}~\bibnamefont
  {Brunet}}, \bibinfo {author} {\bibfnamefont {J.-L.}\ \bibnamefont {Thomas}},
  \bibinfo {author} {\bibfnamefont {R.}~\bibnamefont {Marchiano}}, \ and\
  \bibinfo {author} {\bibfnamefont {F.}~\bibnamefont {Coulouvrat}},\
  }\href@noop {} {\bibfield  {journal} {\bibinfo  {journal} {New J. Phys.}\
  }\textbf {\bibinfo {volume} {11}},\ \bibinfo {pages} {013002} (\bibinfo
  {year} {2009})}\BibitemShut {NoStop}%
\bibitem [{\citenamefont {Paterson}(1996)}]{paterson1996}%
  \BibitemOpen
  \bibfield  {author} {\bibinfo {author} {\bibfnamefont {C.}~\bibnamefont
  {Paterson}},\ }\href@noop {} {\bibfield  {journal} {\bibinfo  {journal} {Opt.
  Commun.}\ }\textbf {\bibinfo {volume} {124}},\ \bibinfo {pages} {121}
  (\bibinfo {year} {1996})}\BibitemShut {NoStop}%
\bibitem [{\citenamefont {Firouzi}\ \emph {et~al.}(2012)\citenamefont
  {Firouzi}, \citenamefont {Cox}, \citenamefont {Treeby},\ and\ \citenamefont
  {Saffari}}]{firouzi2012}%
  \BibitemOpen
  \bibfield  {author} {\bibinfo {author} {\bibfnamefont {K.}~\bibnamefont
  {Firouzi}}, \bibinfo {author} {\bibfnamefont {B.}~\bibnamefont {Cox}},
  \bibinfo {author} {\bibfnamefont {B.}~\bibnamefont {Treeby}}, \ and\ \bibinfo
  {author} {\bibfnamefont {N.}~\bibnamefont {Saffari}},\ }\href@noop {}
  {\bibfield  {journal} {\bibinfo  {journal} {J. Acoust. Soc. Am.}\ }\textbf
  {\bibinfo {volume} {132}},\ \bibinfo {pages} {1271} (\bibinfo {year}
  {2012})}\BibitemShut {NoStop}%
\bibitem [{\citenamefont {Abramowich}\ and\ \citenamefont
  {Stegun}(1972)}]{Abramowich}%
  \BibitemOpen
  \bibfield  {author} {\bibinfo {author} {\bibfnamefont {M.}~\bibnamefont
  {Abramowich}}\ and\ \bibinfo {author} {\bibfnamefont {I.~A.}\ \bibnamefont
  {Stegun}},\ }\href@noop {} {\emph {\bibinfo {title} {Handbook of Mathematical
  functions with formulas, graphs and mathematical tables}}}\ (\bibinfo
  {publisher} {Dover Publications, Inc. New York},\ \bibinfo {year}
  {1972})\BibitemShut {NoStop}%
\end{thebibliography}%
%

\end{document}